\documentclass[cpp,fleqn]{w-art}
\usepackage{times}
\usepackage{w-thm}
\usepackage[]{graphicx}

\begin{document}

\DOIsuffix{theDOIsuffix}
\Volume{}
\Issue{}
\Copyrightissue{}
\Month{}
\Year{}

\Receiveddate{date?}
\Reviseddate{date?}
\Accepteddate{date?}
\Dateposted{date?}


\title[EOS for dense hydrogen]{Equation of state for dense hydrogen and plasma
phase transition}

\author[]{B.~Holst\footnote{Corresponding author: e-mail: {\sf
      bastian.holst@uni-rostock.de}, Phone: +49\,381\,498\,6919,
    Fax: +49\,381\,498\,6912}\inst{1}}
\address[\inst{1}]{Universit\"at Rostock, Institut f\"ur Physik, D-18051
  Rostock, Germany}
\author[]{N.~Nettelmann \inst{1}}
\author[]{R.~Redmer \inst{1}}

\begin{abstract}
  We calculate the equation of state of dense hydrogen within the chemical
  picture. Fluid variational theory is generalized for a multi-component
  system of molecules, atoms, electrons, and protons. Chemical equilibrium is
  supposed for the reactions dissociation and ionization. We identify the
  region of thermodynamic instability which is related to the plasma phase
  transition. The reflectivity is calculated along the Hugoniot curve and
  compared with experimental results. The equation-of-state data is used to
  calculate the pressure and temperature profiles for the interior of
  Jupiter. 
\end{abstract}
\maketitle                  

\section{Introduction}

The equation of state (EOS) of hydrogen and helium at high pressures is of
great relevance for models of the interior of giant planets and other
astrophysical objects as well as for inertial confinement fusion experiments.
For detailed calculations accurate knowledge of the EOS over a wide range of
densities and temperatures is needed.  Especially, in the range of
\textit{warm dense matter} with high densities characteristic for condensed
matter and at temperatures of a few eV the EOS is crucial for modelling giant
planets. This region is challenging for many-particle theory because strong
correlations dominate the physical behavior. Progress in shock-wave
experimental technique has allowed to study this region only recently. 

To probe the EOS, experimental investigations were performed statically with
diamond anvil cells or dynamically by using shock waves, see~\cite{Nellis06}
for a recent review. The experimental data indicate that a nonmetal-to-metal
transition occurs at about 1~Mbar which is identified by a strong increase of
the conductivity~\cite{weir} and reflectivity~\cite{Celliers00}. Some
theoretical models yield a thermodynamic instability in this transition
region, the \textit{plasma phase transition}
(PPT)~\cite{SC1,SC2,EbelNorm,BHRO,FIL06}, which would strongly affect models
for planetary interiors and the evolution of giant
planets~\cite{MIT-Planets,CSHL,Planets-PPT}. After a long period of
controversial discussions, new results of shock wave experiments on deuterium
support the existence of such a PPT~\cite{Fortov}. 
This fundamental problem of high-pressure physics will also be studied with 
the FAIR facility at GSI Darmstadt within the LAPLAS project, 
see~\cite{Tahir,Tahir-PNP12}.

In this paper we present new results for the EOS of dense hydrogen within the
chemical picture. We treat the reactions pressure dissociation and ionization
self-consistently via respective mass action laws. We identify the region of
thermodynamic instability and calculate the phase diagram as well as the
reflectivity in order to verify the corresponding nonmetal-to-metal
transition. The EOS data is used to model the interior of Jupiter within a
three-layer model. The agreement with astrophysical constraints such as the
core mass and the fraction of heavier elements can serve as an additional test
of the theoretical EOS.

\section{Equation of state for dense hydrogen}

Warm dense hydrogen is considered as a partially ionized plasma in the
chemical picture. A mixture of a neutral component (atoms and molecules) and a
plasma component (electrons and protons) is in chemical equilibrium with
respect to dissociation and ionization.  The EOS is derived from an expression
for the free energy of the neutral ($F_0$) and charged particles ($F_\pm$),
see~\cite{SFB-CPP,SCCS05}:
\begin{equation}
F(T,V,N)=F_0+F_\pm+F_{pol}.
\end{equation}
The first two terms consist of ideal and interaction contributions and can be
written as $F_0=F_0^{id}+F_0^{int}$ and $F_\pm=F_\pm^{id}+F_\pm^{int}$.
$F_{pol}$ contains interaction terms between charged and neutral components
caused by polarization~\cite{PolRR}.

Applying fluid variational theory (FVT), the EOS is determined by calculating
the free energy $F_0^{int}(T,V,N)$ via the Gibbs-Bogolyubov
inequality~\cite{RRY}. This method has been generalized to two-component
systems with a reaction~\cite{FVT1,FVT2,N2} so that also molecular systems at
high pressure can be treated where pressure dissociation occurs, e.g.\ H$_2
\rightleftharpoons$ 2H for hydrogen.  In chemical equilibrium, $\mu_{\rm
  H_2}=2\mu_{\rm H}$ is fulfilled, and the number of atoms and molecules can
be determinded self-consistently via the chemical potentials $\mu_c=(\partial
F/\partial N_c)_T$.  The effective interactions between the neutral species
are modeled by exp-6 potentials, and the free energy of a multi-component
reference system of hard spheres has to be known; for details,
see~\cite{FVT1,FVT2,PCCP05}. 

The charged component is treated by using efficient Pad\'{e} approximations
for the free energy developed by Chabrier and Potekhin~\cite{Potek}. The
coupling with the neutral component occurs via the ionization equilibrium,
H$\rightleftharpoons$e+p. In chemical equilibrium, the relation $\mu_{\rm
  H}=\mu_{\rm e}+\mu_{\rm p}$ determines the degree of ionization.  

\begin{figure}[ht]
\begin{center}
\includegraphics[width=0.5 \textwidth]{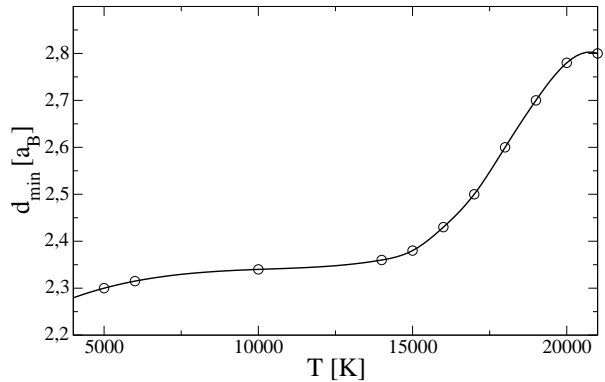}
\end{center}
\caption{Minimum diameter for expanded particles (atoms, molecules) introduced
  within the reduced volume concept. \label{fig:dmin}}
\end{figure}

Since atoms and molecules are particles of finite size there is an additional
interaction between the charged component and the neutral fluid. According to
the concept of reduced volume, point-like particles cannot penetrate into the
volume occupied by atoms and molecules. This leads to a correction in the
description of the ideal gas of the charged
component~\cite{McLellan56,Kahlbaum92} so that the ideal free energy of
protons and electrons $F_\pm^{id}$ is dependent on the reduced volume
$V^*=V\cdot(1-\eta)$,
\begin{equation}
F_\pm^{id}(T,V^*,N)=N_\pm k_B T \cdot f_\pm^{id,*}, 
\end{equation}
where $\eta$ is the ratio of the volume which cannot be penetrated by
point-like particles to the total volume.  It is derived from hard sphere
diameters obtained within the FVT self-consistently. The free energy density
$f_\pm^{id,*}$ is given by Fermi integrals which take into account quantum
effects. In order to avoid an intersection of pressure isotherms, which is
important for modelling planetary interiors, a minimum diameter $d_\text{min}$
has been introduced.  It was determined starting at low temperatures where it 
remains almost constant up to 15.000~K, then it increases up to 20.000~K and
remains constant again for higher temperatures, see
Fig.~\ref{fig:dmin}.  These values are in the range of the results for the
diameter of the hydrogen atom derived from the confined atom
model~\cite{confatom}.

Consequently, the reduced volume concept changes the chemical potential of
each component drastically at higher densities and results in pressure
ionization. This is due to the fact that additional terms appear in the
chemical potential, which is the particle number derivative of the free
energy, and thermodynamic functions of degenerate plasmas are very sensitive
to changes in density.

This current model FVT$^+$ includes all interaction
contributions to the chemical potentials, thus being a generalization
of earlier work~\cite{PCCP05} where only ideal plasma contributions have been
treated (FVT$^+_{\rm id}$).   

\begin{figure}[htb]
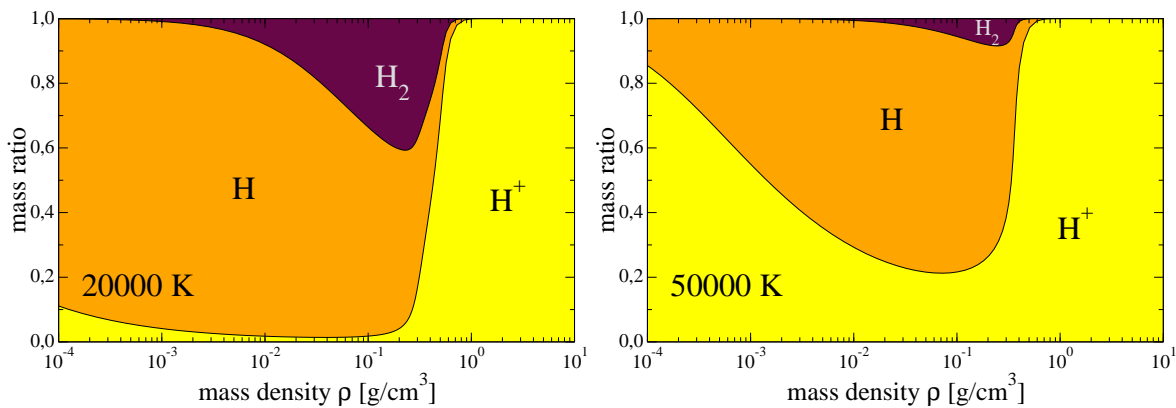

\begin{center}
\includegraphics[width=0.49 \textwidth]{bilder/comp020}
\includegraphics[width=0.49 \textwidth]{bilder/comp050}
\end{center}
\caption{Composition of dense hydrogen for 20.000~K (left) and 50.000~K
  (right). \label{fig:compo}}
\end{figure}

In Fig.~\ref{fig:compo} the composition of hydrogen derived from the present
approach is shown for two temperatures. Hydrogen is an atomic gas at low
temperatures (left) and low densities. With increasing densities molecules are
formed due to the mass action law. Pressure dissociation and ionization can be
observed in the high-density region. The nonideality corrections to the free
energy force a transition from a molecular fluid to a fully ionized plasma. At
higher temperatures (right) the formation of molecules is suppressed and
pressure ionization becomes the dominating process. At low densities and high
temperatures a fully ionized plasma is produced due to thermal ionization.

\begin{figure}[htb]
\begin{center}
\includegraphics[width=0.5 \textwidth]{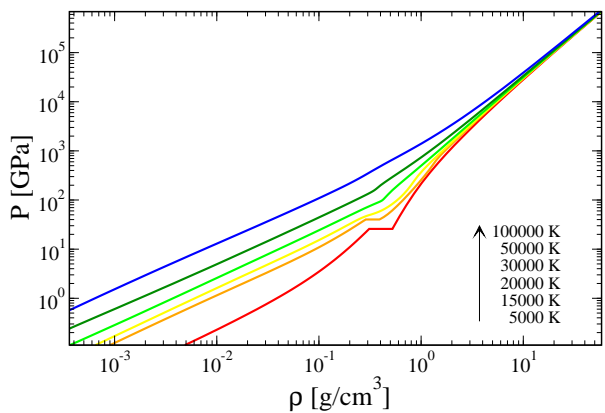}
\end{center}
\caption{Pressure isotherms for dense hydrogen. \label{fig:Piso}}
\end{figure}

We show pressure isotherms over a wide range of temperatures and densities in
Fig.~\ref{fig:Piso}. At low densities the system behaves like a neutral fluid.
Between densities of 10$^{-3}$ g/cm$^3$ and 10$^{-1}$ g/cm$^3$
nonideality corrections to the free energy of atoms and molecules lead to a
nonlinear behavior of the isotherms.  For still higher densities a phase
transition occurs which is treated by a Maxwell construction.  The
thermodynamic instability vanishes with increasing temperatures, and the
critical point is located at 16.800~K, 0.35~g/cm$^3$, and 45~GPa.

\begin{figure}[htb]
\begin{center}
\includegraphics[width=0.5 \textwidth]{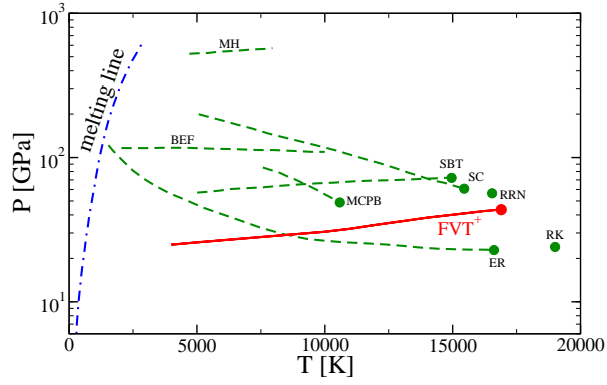}
\end{center}
\caption{Phase diagram for dense hydrogen. Present results of the FVT$^+$
  (red) are compared with other predictions for the PPT:
  SC~\protect\cite{SC1,SC2}, RK~\protect\cite{Robnik}, 
  MH~\protect\cite{MH}, ER~\protect\cite{EbelRich}, 
  SBT~\protect\cite{SBT}, RRN~\protect\cite{RRN}, 
  BEF~\protect\cite{BEF}, MCPB~\protect\cite{MCPB}.  
\label{fig:PT}}
\end{figure}

The critical point and the related coexistence line are shown in
Fig.~\ref{fig:PT} and compared with results of other EOS. The critical point
itself lies within the range of other predictions, whereas the coexistence
line is lower than most of the other results. For a comparison of data
concerning the PPT, see Table~\ref{tab:cp}.

New shock-wave experiments~\cite{Fortov} imply that a PPT occurs in deuterium
at densities of 1.5 g/cm$^3$ and a coexistence pressure of about 1 megabar.
Each of these values is twice as high as evaluated in the recent model.

\begin{table}[htb] 
\begin{footnotesize}
\begin{center}
\begin{tabular}{ccclll} \hline\hline

$T_c$  & $p_c$  & $\rho_c$   & Method & Authors & Reference \\
(10$^3$ K) & (GPa) & (g/cm$^3$) & & & \\ \hline

12.6  & 95     & 0.95   & PIP    & Ebeling/S\"andig (1973) & 
 \protect\cite{EbelSand} \\
19    & 24     & 0.14   & PIP    & Robnik/Kundt (1983) & 
 \protect\cite{Robnik}\\ 
16.5  & 22.8   & 0.13   & PIP    & Ebeling/Richert (1985) & 
 \protect\cite{EbelRich} \\
16.5  & 95     & 0.43   & PIP    & Haronska {\it et al.} (1987) & 
 \protect\cite{Harry} \\
15    & 64.6   & 0.36   & PIP    & Saumon/Chabrier (1991) & 
 \protect\cite{SC1}\\
15.3  & 61.4   & 0.35   & PIP    & Saumon/Chabrier (1992) & 
 \protect\cite{SC2}\\
14.9 & 72.3    & 0.29   & PIP    & Schlanges {\it et al.} (1995) & 
 \protect\cite{SBT}\\ 
16.5  & 57     & 0.42   & PIP    & Reinholz {\it et al.} (1995) & 
 \protect\cite{RRN}\\
11    & 55     & 0.25   & PIMC   & Magro {\it et al.} (1996) & 
 \protect\cite{MCPB}\\
20.9  & 0.3    & 0.002  &  & Kitamura/Ichimaru (1998) & 
 \protect\cite{Ichi}\\
16.8  & 45     & 0.35   & PIP    & present FVT$^+$ & \\
\hline\hline

\end{tabular}
\end{center}
\end{footnotesize}
\caption{\label{tab:cp}Theoretical results for the critical point of the 
  hypothetical plasma~phase transition (PPT) in hydrogen which was
  predicted by Zeldovich and Landau~\protect\cite{Landau} and Norman and
  Starostin~\protect\cite{Norman}. } 
\end{table}


\section{Conductivity and reflectivity}

The PPT is an instability driven by the nonmetal-to-metal transition (pressure
ionization). We calculate the electrical conductivity as well as the
reflectivity by applying the COMPTRA04 program
package~\cite{Comptra04,www-comptra} in order to locate this transition in the
density-temperature plane.

Optical properties are calculated within the Drude model.  The reflectivity
$R(\omega)$ is given in the long-wavelength limit via the dielectric function
$\varepsilon(\omega)$ which is determined by a dynamic collision frequency
$\nu(\omega)$ or, alternatively, by the dynamic conductivity
$\sigma(\omega)$~\cite{collision}: 
\begin{eqnarray}
   R(\omega) &=& 
 \left|\frac{ \sqrt{\varepsilon(\omega)}-1}{\sqrt{\varepsilon(\omega)}-1} 
 \right|^2,\\
   \varepsilon(\omega)&=& 1-\frac{\omega_{pl}^2}
 {\omega\left[\omega+\text{i}\nu(\omega)\right]}
 = 1+\frac{\text{i}}{\varepsilon_0\omega}\sigma(\omega) ,\\
   \sigma(\omega)&=& \sigma(0)\left[
 1-\frac{\text{i}\omega}{\varepsilon_0\omega_{\text{pl}}^2}
 \sigma(0) \right]^{-1}.
\end{eqnarray}
$\omega_{\text{pl}}=\sqrt{n_e e^2/(\varepsilon_0 m_e)}$ is the plasma frequency 
of the electrons.

\begin{figure}[htb]
\begin{center}
\includegraphics[width=0.5 \textwidth]{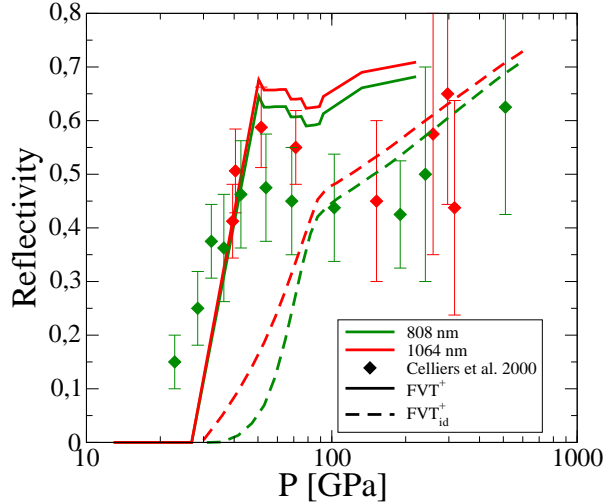}
\end{center}
\caption{Reflectivity of dense hydrogen within the models FVT$^+$ and
  FVT$^+_{\rm id}$ along the Hugoniot curve in comparison
  with experiments~\protect\cite{Celliers00}.  \label{fig:reflex}}
\end{figure}

The reflectivity was determined along the Hugoniot curve and is compared with
experimental results~\cite{Celliers00} and those of the earlier model
FVT$^+_{\rm id}$~\cite{SCCM} in Fig.~\ref{fig:reflex}.  The results of the
current model show a much better agreement with the experiment. The
characteristic and abrupt rise with increasing pressure was reproduced more
accurately. This drastic increase appears due to pressure ionization in the
vicinity of the criotical point of the PPT. As a result, the reflectivity
advances from very low values to metallic-like ones almost instantly.

\section{Planetary interiors}

Modelling the interiors of giant planets and comparison with their
observational parameters offers an alternative tool besides laboratory
experiments of probing the EOS of the components the planets are predominantly
made of. Giant planets such as Jupiter and Saturn consist mainly of hydrogen
and, in decreasing order, of helium, water and rocks, covering a wide range of
pressures and temperatures.  Independently from the H-EOS used for modelling,
the simplest interior structure that is compatible with the observational
constraints requires at least three homogenous layers with a transition from a
cold molecular fluid in the outer envelope to a pressure ionized plasma in the
deep interior and a dense solid core of ices and rocks. A solid core may be
explained as a result of the formation process and the seperation into two
fluid envelopes with different particle abundances by an existence of a PPT as
provided by the FVT$^+$ EOS. The constraining observational parameters are the
total mass of the planet $M$, its equatorial radius $R_{eq}$, the temperature
$T$ at the outer boundary, the average helium content $\bar{Y}$, the period of
rotation $\omega$ and the gravitational moments $J_2, J_4, J_6$. From
measurements of the luminosity it has been argued \cite{Hubbard68} that the
temperature profile should be adiabatic.  For a given EOS, the interior
profiles of pressure $P$ and density $\rho$ are calculated by integration of
the equation of hydrostatic equilibrium
\begin{equation}
	\label{eq_hydrostatGG}
	\frac{1}{\rho(r,\theta)} \nabla_{\vec{r}} P(r,\theta) = 
	\nabla_{\vec{r}}
\left(G\int_{dV}d^3r'\frac{\rho(r,\theta)}{|\vec{r}-\vec{r}'|} 
		+  \frac{1}{2}\omega^2 r^2\sin\theta^2\right)
\end{equation}
along an isentrope defined by the outer boundary. The first term on the right
hand side of eq. (\ref{eq_hydrostatGG}) is the gravitational potential and the
second term the centrifugal potential assuming axialsymmetric rotation. We
apply the \textsl{theory of figures} \cite{ZharkovTrubitsyn} up to third order
to solve this equation and to calculate the gravitational moments. They are
defined as the coefficients of the expansion of the gravitational potential
into Legendre polynomials, taken at the outer boundary. Being integrals of the
density distribution weighted by some power of the radius, they are very
sensitive with respect to the amount and distribution of helium and heavier
elements within the planet.

In accordance with previous calculations \cite{SC92,TGs}, mixtures of hydrogen
with helium and heavier elements have been derived from the EOS of the pure
materials via the additive volume rule. It states that the entropy of mixing
can be neglected.

\begin{figure}[htb]
\begin{center}
\rotatebox{270}{\includegraphics[width=6cm]{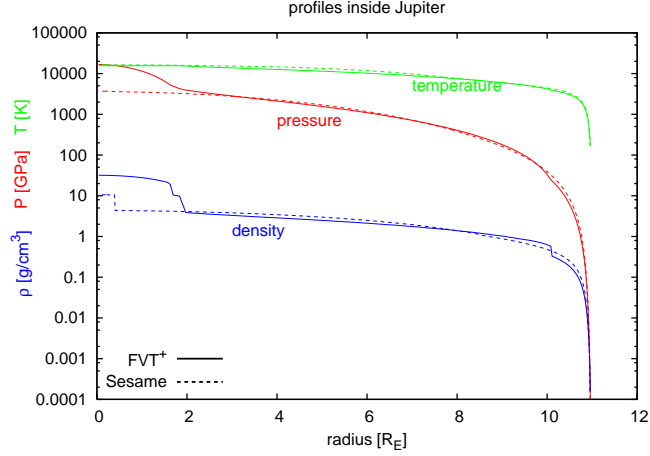}}
\end{center}
\caption{\label{fig-profileJup} 
  Profiles of temperature, density, pressure
  along the radius within Jupiter using two different H-EOS, FVT$^+$ (solid)
  and Sesame 5251 (dashed).}
\end{figure}

Assuming a three-layer structure, we present results for Jupiter for the
profiles of temperature, density, and pressure along the radius in
Fig.~\ref{fig-profileJup} using two different H-EOS, the standard Sesame table
5251 for hydrogen~\cite{Sesame} and the FVT$^+$ model presented above.

The profiles of temperature appear very similar, meaning a small uncertainty
about the real profiles. Contrary, the density and pressure profiles exhibit
more differences and require some explanation. In the fluid part of Jupiter,
the presence of a PPT leads to a jump in density between the envelopes. Since
the gravitational moments as integrals over the density have to be the same
for both H-EOS, the density profile of a H-EOS with PPT has to be smaller in
the outer envelope and larger in the inner envelope. The different size and
composition of the core for these specific H-EOS are a consequence of their
different compressibility in the regime of pressure ionization at about
1~Mbar, where the gravitational moments are most sensitive to the density
distribution. 

In case of a stiff H-EOS like Sesame, a larger amount of heavy elements is
needed in the two fluid envelopes to compensate for the smaller hydrogen
density at a given pressure.  As a result, this material is added to the
well-known density-pressure relation of degenerate electrons in the deep
interior, leaving less material for the core.  Thus, in case of the
Sesame-EOS, the amount of heavy elements becomes with 10\% very large and an
unlikely solution with a very small core of light material (e.g.\ water) can
be found. 

In case of the FVT$^+$ EOS which is more compressible than the
Sesame EOS at about 1~Mbar, the helium content is below the value of 27.5\%
for the protosolar cloud in order to reproduce the lowest gravitational moment
$J_2$. Furthermore, the next gravitational moment $J_4$ cannot be reproduced
correctly because the transition to the metallic envelope occurs already at
about 90\% of the radius and, thus, at too low densities. For opposite
reasons, both the Sesame and FVT$^+$ EOS applied in a three-layer model of
Jupiter are not compatible with {\it all} of the observational
constraints. While Sesame is probably too stiff, the FVT$^+$ model is likely
too soft in the WDM region at about 1~Mbar.

\section{Conclusions}

In this paper, we have extended the earlier chemical model FVT$^+_{id}$ to
calculate the EOS of dense hydrogen. The current model FVT$^+$ includes
nonideality corrections to the free energy of each commponent of the partially
ionized plasma. We have shown results for the composition and the
thermodynamic properties of dense hydrogen. The PPT was located in the phase
diagram, its critical point coincides with earlier results. Furthermore, we
have determined optical properties such as reflectivity and conductivity,
within linear response theory using the program package COMPTRA04.  The
calculated reflectivity along the experimental Hugoniot curve shows a good
agreement with the experiments. However, application of the FVT$^+$ EOS to the
interior structure of Jupiter indicates that the behavior at about 1~Mbar is
probably too soft. The same conclusion can be drawn from a comparison with
shock-wave experiments that indicate the existence of a PPT~\cite{Fortov}.
FVT$^+$ predicts the PPT at too low pressures as well as at too low densities.
Further efforts to solve this problem, especially concerning the reduced
volume concept, are necessary.

\begin{acknowledgement}
  We thank P.~M.~Celliers, W.~Ebeling, V.~E.~Fortov, V.~K.~Gryaznov,
  W.-D.~Kraeft, and G.~R\"opke for stimulating discussions.  This work was
  supported by the DFG within the SFB 652 Strongly Correlated Matter in
  Radiation Fields and the GRK 567 Strongly Correlated Many Particle Systems.
\end{acknowledgement}


\end{document}